\documentclass[english,aps,preprint]{revtex4}
\usepackage[T1]{fontenc}
\usepackage[latin9]{inputenc}
\usepackage{float}
\usepackage{graphicx}
\usepackage{esint}

\makeatletter
\@ifundefined{textcolor}{}
{%
 \definecolor{BLACK}{gray}{0}
 \definecolor{WHITE}{gray}{1}
 \definecolor{RED}{rgb}{1,0,0}
 \definecolor{GREEN}{rgb}{0,1,0}
 \definecolor{BLUE}{rgb}{0,0,1}
 \definecolor{CYAN}{cmyk}{1,0,0,0}
 \definecolor{MAGENTA}{cmyk}{0,1,0,0}
 \definecolor{YELLOW}{cmyk}{0,0,1,0}
}

\makeatother

\usepackage{babel}
\begin{document}

\title{Inertia and Prediction in the Response to External Perturbation of
Noisy Variables}

\author{Nash Rochman}
\email{nashdeltarochman@gmail.com}

\affiliation{Department of Chemical and Biomolecular Engineering, The Johns Hopkins
University }

\author{Sean X. Sun}
\email{ssun@jhu.edu}

\affiliation{Department of Mechanical Engineering and Biomedical Engineering,\\
The Johns Hopkins University, Baltimore MD 21218}
\begin{abstract}
For most stochastic dynamical systems, variables which are tightly
regulated tend to respond slowly to external changes. This idea is
often discussed for applicable systems, within a linear response regime,
through the Fluctuation Dissipation Theorem (FDT). In a previous paper,
we proposed a phenomenological model for the response of the cell
cycle duration distribution to environmental changes which correlated
the width of this distribution to response efficiency when FDT was
not applicable. Here we emphasize how that model may be used to illustrate
this general principle, that the stochasticity of a variable while
inversely proportional to stability is often directly proportional
to lability. Comparisons are made between this discrete-time model
and the simple harmonic oscillator. We then consider a simple continuous
dynamical system, the ``Active Oscillator'', which illustrates this
principle in another fashion.
\end{abstract}
\maketitle

\section{introduction}

Speed and precision are qualities which often run in opposition to
one another. Within stochastically driven systems noisy variables
are often a source of systematic instability, negatively impacting
predictability while possibly improving adaptability. This phenomenon
is well illustrated by the Fluctuation Dissipation Theorem (FDT).
In the early 20th century a series of observations\cite{johnson1928thermal}
regarding thermal noise in conductors lead to the development of FDT
relating the stochastic fluctuations of an ensemble at equilibrium
around its mean to the relaxation rate of that ensemble when exterior
conditions (e.g. temperature) are modulated\cite{nyquist1928thermal,callen1951irreversibility}.
For example, we may consider a univariate system described by the
Hamiltonian $H(x)$ and the corresponding Gibbs measure $P(x)=exp(-\beta H(x))/\int dx'exp(-\beta H(x'))$.
We perturb the system at time zero such that $H(x,t)=H(x)+f_{0}x$
and further restrict the size of this perturbation to satisfy the
approximation $exp(-\beta f_{0}x)\approx1-\beta f_{0}x$. In this
case, FDT yields the following result:

\begin{equation}
\left\langle x(t)\right\rangle =\left\langle x(\infty)\right\rangle +\left(\left\langle x(0)\right\rangle -\left\langle x(\infty)\right\rangle \right)A(t)
\end{equation}
where $A(t)\equiv\left\langle \left(x(t)-\left\langle x(0)\right\rangle \right)\left(x(0)-\left\langle x(0)\right\rangle \right)\right\rangle /\left\langle x(0)x(0)\right\rangle _{t=0}$
is the autocorrelation function. With this result we find a relationship
between the relaxation of the mean of the distribution after the perturbation
and the autocorrelation of the system at equilibrium. A shorter correlation
at equilibrium (less ``memory'' in the system) yields a faster rate
of response to perturbation. Furthermore, a noisier distribution often
generates a shorter correlation function making the connection between
speed and precision. While generalizations of FDT have been realized
for non-equilibrium systems\cite{verley2011modified}, not all systems
which show a correlation between response dynamics and distribution
width are adequately described this way. For a previous problem\cite{rochman2016grow},
we constructed a generalized two-state model transition probability
balancing self-similarity and adaptation. Below we revisit this model
in a general setting and compare it to two related dynamical systems.

\section{Modeling}

We consider the following transition rule for moving from state $y$
in the current timestep to state $x$ in the subsequent timestep:

\begin{equation}
M(y\rightarrow x)=Aexp\left(-\left(y+x\right){}^{2}\right)exp\left(-\left(\frac{1-\xi}{\xi}\right)\left(y-x\right){}^{2}\right)
\end{equation}
where $0<\xi<1$ and $A$ is the normalization constant. The first
term represents the optimization term, a tendency to move away from
the current state towards the ``optimal'' state (zero) and the second
is the inertial term: a tendency to stay in the current state. Rewriting
as a single Gaussian function yields:

\begin{equation}
M(y\rightarrow x)=\sqrt{\frac{1}{\pi\xi}}exp\left(-\left(x-y\left(1-2\xi\right)\right)^{2}/\xi\right)
\end{equation}
We can write down the mean of the distribution for the $n^{th}$ timestep
as follows:

\begin{equation}
\mu_{n}=\mu+\Pi\left(\mu_{n-1}-\mu\right),\,\Pi\equiv1-2\xi
\end{equation}
and the stationary distribution ($P^{*}(y)$ such that $\int MP^{*}(y)dy=P^{*}(x)$):

\begin{equation}
P^{*}(y)=Aexp\left(-4\left(1-\xi\right)y^{2}\right)
\end{equation}
Now we can examine the ``relaxation time'' defined here to be the
number of timesteps it takes for the mean of the distribution to fall
within $e^{-3}\sim5\%$ of the starting value $\Delta$:

\begin{equation}
\left|\left(1-2\xi\right)^{N}\Delta\right|=Err=\Delta e^{-3}
\end{equation}
We may scale the error by the initial mean displacement $\Delta$
to yield:

\begin{equation}
N=\frac{-3}{ln\left(\left|1-2\xi\right|\right)}
\end{equation}
where in principle $N$ must really be the ceiling integer for this
value. Similarly, we examine how the standard deviation of the stationary
distribution varies with $\xi$:

\begin{equation}
\sigma\left(\xi\right)=\sqrt{\frac{1}{2\left(4\left(1-\xi\right)\right)}}=\sqrt{\frac{1}{8\left(1-\xi\right)}}
\end{equation}
We may now look at how the behavior varies across the possible value
of $\xi$:

\begin{figure}[H]
\noindent \begin{centering}
\includegraphics[scale=0.5]{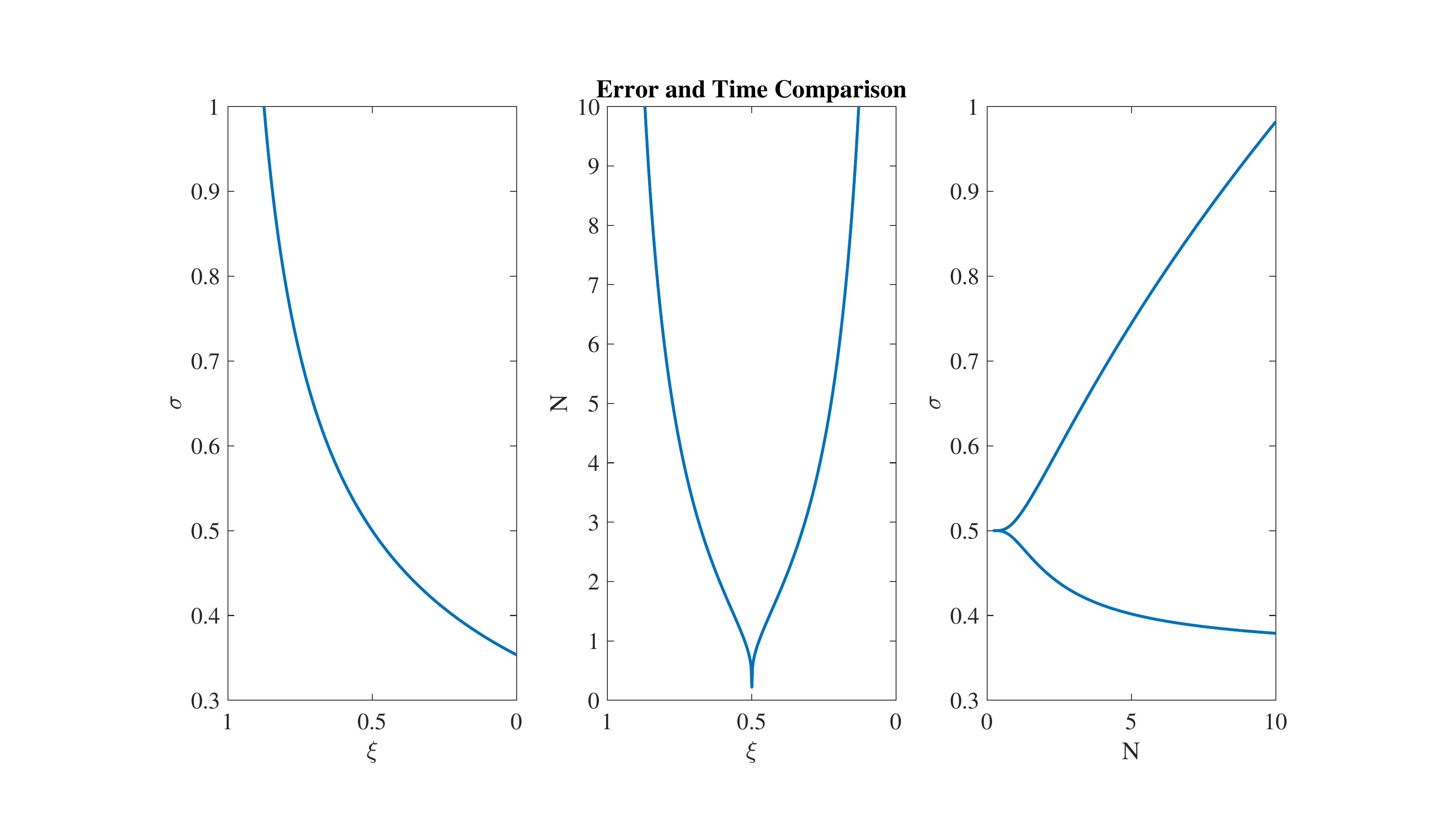}
\par\end{centering}
\caption{Trajectories for all possible values of $\xi$: as the parameter tends
to zero and the weight on the self-similar or inertial term diverges,
the spread of the final distribution shrinks, but the relaxation time
diverges. }
\end{figure}
We find that increasing the weight on the self-similar or inertial
term (decreasing $\xi$) while shrinking the spread of the stationary
distribution (the ``error'') increases the relaxation time. A good
compromise is selecting $\xi=0.5$ where the relaxation time is minimized
and the error is still low ($1/2$). In other words, decreasing the
noise in the system, while improving precision, compromises adaptability
and the relaxation time.

We may compare the model presented above to the damped harmonic oscillator:

\begin{equation}
\frac{d^{2}x}{dt^{2}}+\beta\frac{dx}{dt}+x=0
\end{equation}
We will consider the initial conditions $x\left(0\right)=1$ and $\frac{dx}{dt}\left(0\right)=0$
to match our work above and additionally introduce the parameter $\lambda$
which varies from zero to one: $\beta=2\left(\frac{\lambda}{1-\lambda}\right)$.
There are three forms for the solutions: overdamped, critically damped,
and underdamped corresponding to real distinct, real repeated, and
imaginary roots respectively. For each case we have the following
solutions. Overdamped:

\begin{equation}
x\left(t\right)=\frac{1}{2}e^{-\frac{\beta}{2}t}\left[\left(1-\sqrt{\frac{\beta^{2}}{\beta^{2}-4}}\right)e^{-\frac{1}{2}t\sqrt{\beta^{2}-4}}+\left(1+\sqrt{\frac{\beta^{2}}{\beta^{2}-4}}\right)e^{\frac{1}{2}t\sqrt{\beta^{2}-4}}\right]
\end{equation}
Critically Damped:

\begin{equation}
x\left(t\right)=\left(1+t\right)e^{-t}
\end{equation}
Underdamped:

\begin{equation}
x\left(t\right)=e^{-\frac{\beta t}{2}}\left[cos\left(\frac{1}{2}\sqrt{\beta^{2}-4}t\right)+\sqrt{\frac{\beta^{2}}{\beta^{2}-4}}sin\left(\frac{1}{2}\sqrt{\beta^{2}-4}t\right)\right]
\end{equation}
We will define the relaxation time as the time when all subsequent
points in the trajectory fall below the threshold value $e^{-3}$.
There isn't really an analogous ``error'' term here since all trajectories
(regardless of parameter value) eventually reach the origin; however,
we can compare the magnitude of the values at some specified time
(i.e. $\left|x\left(1\right)\right|$) for a similar measure. We'll
label time $s$ and ``error'' $\varepsilon$:

\begin{figure}[H]
\noindent \begin{centering}
\includegraphics[scale=0.5]{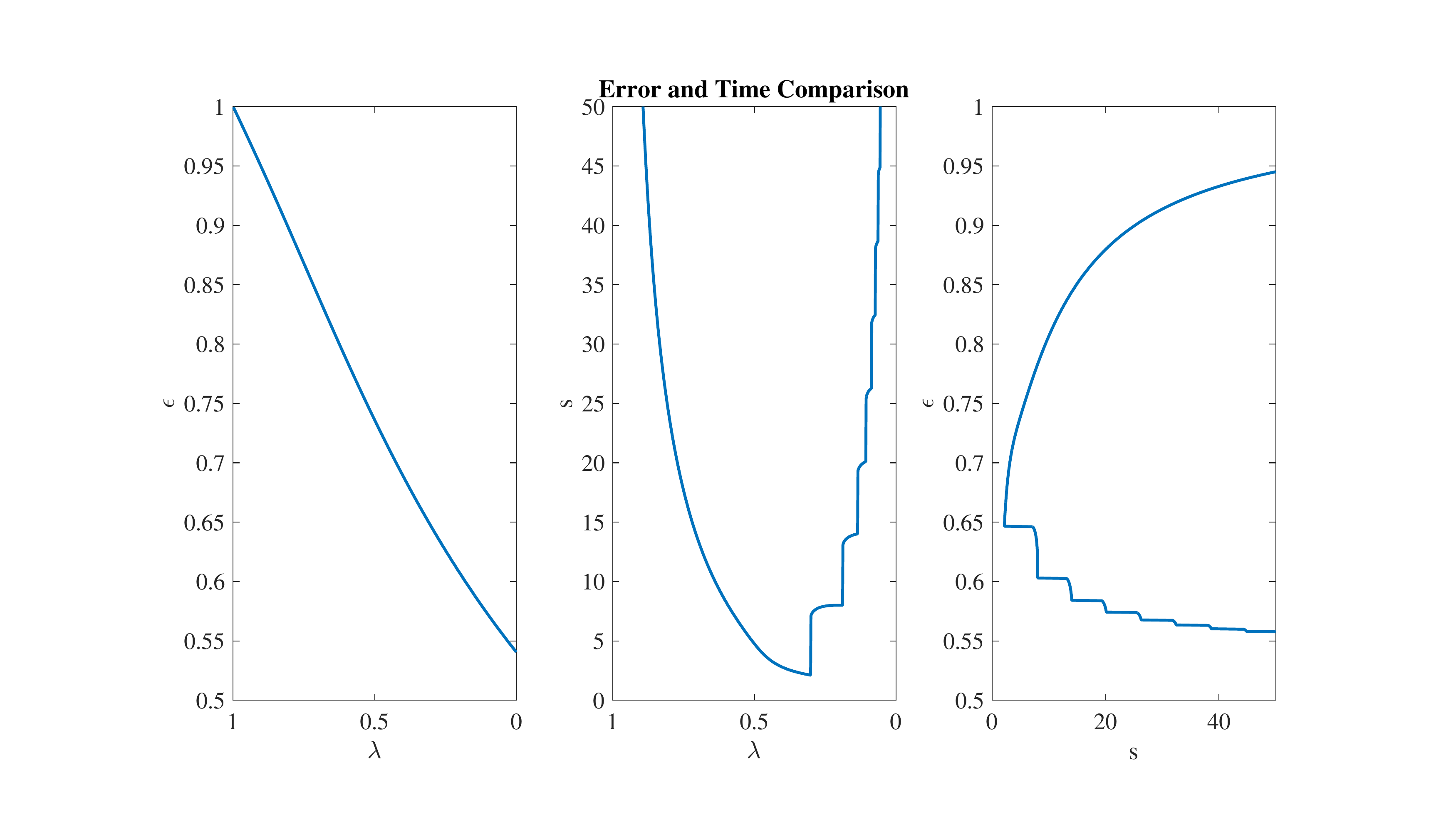}
\par\end{centering}
\caption{Trajectories for all possible values of $\lambda$. We observe the
trend that as the parameter tends to zero the error shrinks but the
relaxation time diverges.}
\end{figure}
We find that the curves qualitatively match those for the discrete-time
model and observe the same trend: as the parameter decreases, the
error decreases but the relaxation time increases. Below are some
representative trajectories from these models for comparison. Parameter
values of $\xi=\lambda=\frac{1}{4},\frac{1}{3},\frac{1}{2},\frac{2}{3}$
and $\frac{3}{4}$ are shown:

\begin{figure}[H]
\noindent \begin{centering}
\includegraphics[scale=0.5]{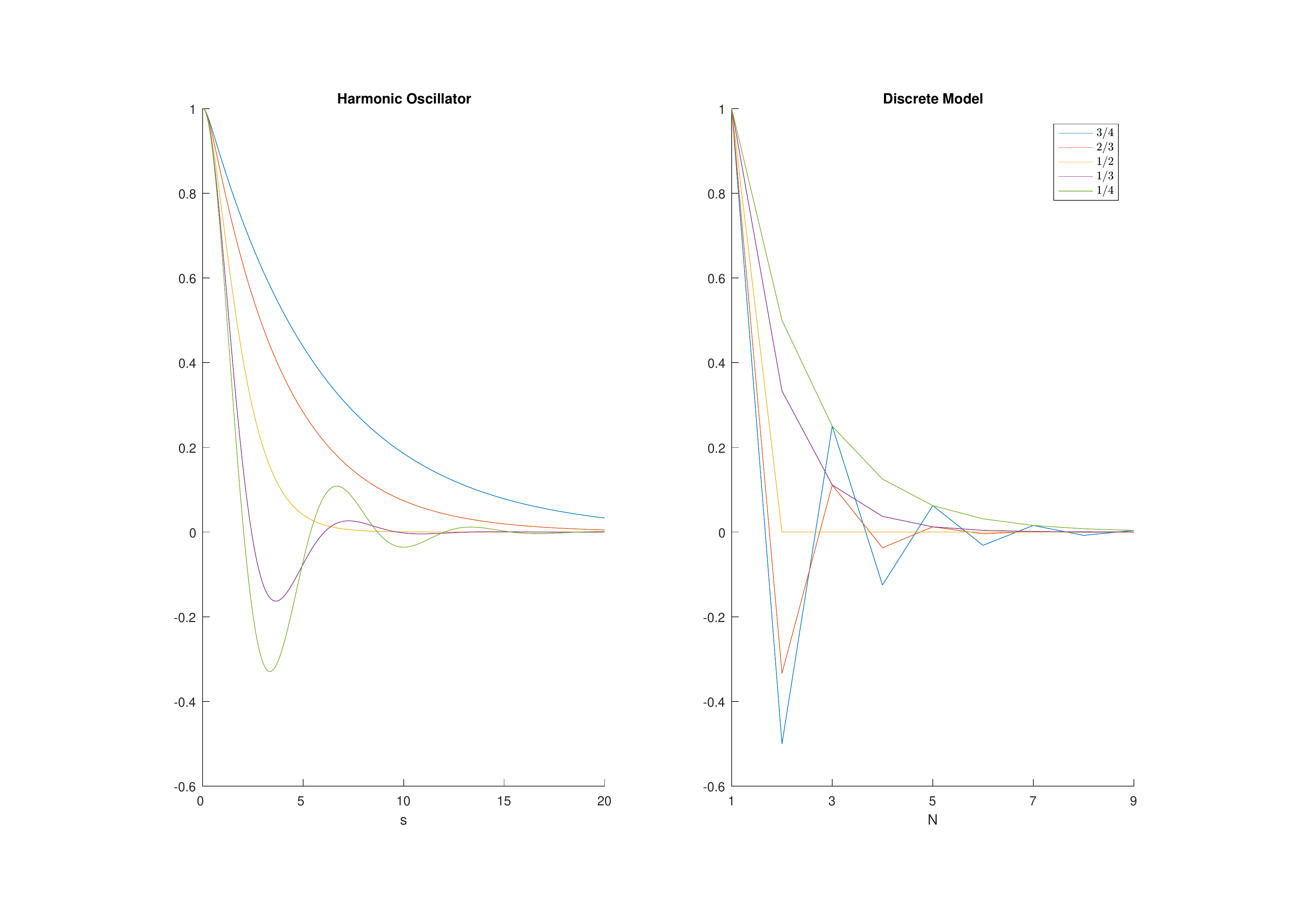}
\par\end{centering}
\caption{Comparing individual trajectories for parameter values $\xi=\lambda=\frac{1}{4},\frac{1}{3},\frac{1}{2},\frac{2}{3}$
and $\frac{3}{4}$  for the discrete model and harmonic oscillator.}
\end{figure}

We would now like to introduce the complementary, ``Active Oscillator''
toy model. Consider an object with an initial value $x=-\Delta$,
zero initial velocity $x'=0$, and a target position of zero. Next
let us define the projected state $x^{p}(t)$, or the value obtained
from a kinematic prediction of the value some time $\tau>0$ into
the future:

\begin{equation}
x^{p}(t)=x(t)+x'(t)\tau+\frac{1}{2}x"(t)\tau^{2}
\end{equation}
Let us consider a set of simple rules. The acceleration maintains
a value of $\alpha$ until the projected state reaches zero at which
time ($t_{1}$) the acceleration is set to zero. The position then
continues to increase until ($t_{1}+t_{2}$) it reaches zero, and
the acceleration is set to $-\alpha$. This state is maintained until
($t_{1}+t_{2}+t_{3}$) when the velocity reaches zero and the trace
either finishes or repeats as discussed below. 

First we may scale the variables by the characteristic length scale
$\Delta$ and timescale $\sqrt{\frac{2\Delta}{\alpha}}$ for convenience.
The projected state reaches zero when:

\begin{equation}
x^{p}(0)=-1+\tau^{2}
\end{equation}
When $\tau\geq1$, we find that the projected state is already positive
and the object doesn't ever move. For $\tau<1$ we find:

\begin{equation}
x^{p}(t_{1})=0=-1+\left(\tau+t_{1}\right)^{2}\Rightarrow t_{1}=1-\tau
\end{equation}
The position reaches zero, $x\left(t_{1}+t\right)=0$, when:

\begin{equation}
x\left(t_{1}+t\right)=-1+\left(1-\tau\right)^{2}+2\left(1-\tau\right)t=0\Rightarrow t_{2}=\frac{2\tau-\tau^{2}}{2\left(1-\tau\right)}
\end{equation}
The velocity reaches zero at time $t_{1}+t_{2}+t_{3}$, where $t_{3}=t_{1}=1-\tau$.
The final position (when the velocity is zero) is given by:

\begin{equation}
X=x\left(t_{1}+t_{2}+t_{3}\right)=2\left(1-\tau\right)^{2}-\left(1-\tau\right)^{2}=\left(1-\tau\right)^{2}
\end{equation}
and the total time it takes to reach this point is:

\begin{equation}
T=t_{1}+t_{2}+t_{3}=2\left(1-\tau\right)+\frac{2\tau-\tau^{2}}{2\left(1-\tau\right)}
\end{equation}
This may be repeated for smaller values of $\tau$ where $\Delta=X$.
The results for the second cycle are (where $X$ is reported positive
for convenience) $X=\left(1-2\tau\right)^{2}$ and $T=2\left(1-2\tau\right)+\frac{2\tau-3\tau^{2}}{2\left(1-2\tau\right)}$
and for the third cycle, $X=\left(2\tau^{2}-4\tau+1\right)^{2}$ and
$T=2\left(2\tau^{2}-4\tau+1\right)+\frac{4\tau^{3}-7\tau^{2}+2\tau}{2\left(2\tau^{2}-4\tau+1\right)}$
. The fourth cycle begins at begins at: $x^{p}(0)=-\left(1-\tau\right)^{2}\left(1-2\tau\right)^{2}\left(2\tau^{2}-4\tau+1\right)^{2}+\tau^{2}\Rightarrow\tau<0.18$.
The results for the first three cycles are shown below: 

\begin{figure}[H]
\noindent \begin{centering}
\includegraphics[scale=0.5]{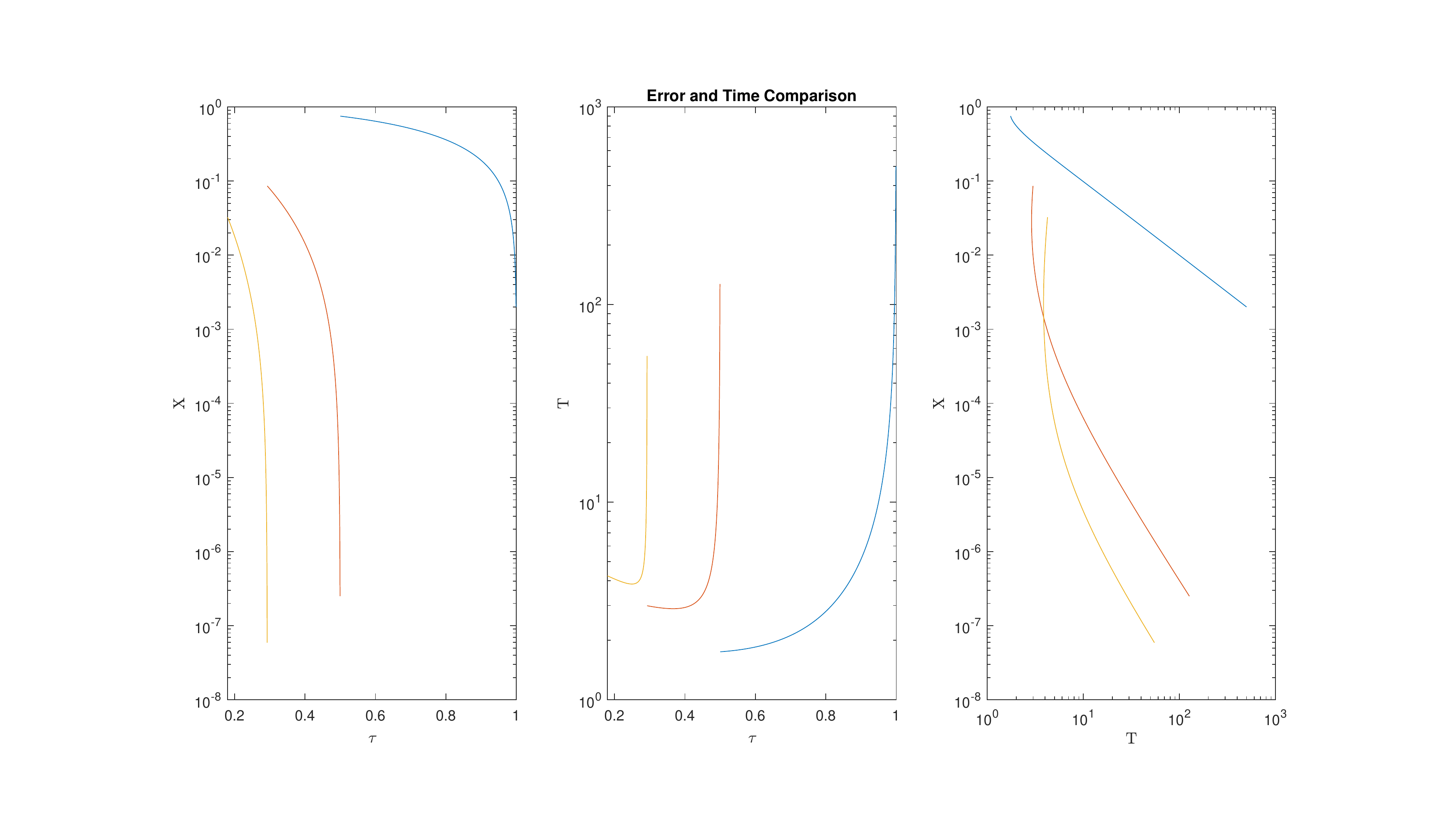}
\par\end{centering}
\caption{The first three cycles for the Active Oscillator corresponding to
values of $\tau$ greater than $\sim0.18$. }
\end{figure}
The behavior for this system is messier than the previous two; however,
within each cycle analogous trends are observed - increasing $\tau$
decreases error and increases relaxation time. The projection length,
$\tau$, plays an analogous role to the weight on the self-similar
term $\xi$ with large values of $\tau$ corresponding to small values
of $\xi$.

\section{discussion}

Tightly regulated variables often respond slowly to external changes.
We compared two models which display this characteristic stemming
from different underlying mechanisms. In the first case, we considered
a two-term transition probability in which the second ``inertial''
term represented a tendency to maintain the current state. The behavior
of this system mirrored that of the harmonic oscillator in many ways.
When the weight, $\xi$, on the inertial term tended from one half
to one, the relaxation time increased and the standard deviation of
the resulting distribution increased. While the actual trajectories
resemble the under-damped case, the relationship between error and
relaxation time is similar to the overdamped oscillator where $\lambda$
is greater than one half. As $\xi$ tends from one half to one, the
standard deviation decreases but the relaxation time still increases
resembling the underdamped oscillator where as $\lambda$ tends to
zero, the first zero crossing occurs earlier (and the ``error''
reported above decreases) but the relaxation time diverges. In the
second model, we examined another oscillator where the acceleration
was partly determined by a kinematic prediction time $\tau$ into
the future. The behavior of this model was more complex, but within
repeated regions we found that as $\tau$ increases, the relaxation
time increases and the error decreases. Both increasing $\tau$ and
decreasing $\xi$ tightens a regulatory constraint and doing so makes
responses slower but errors smaller.

\section{references}

\bibliographystyle{plain}

\begin{thebibliography}{1}

\bibitem{callen1951irreversibility}
Herbert~B Callen and Theodore~A Welton.
\newblock Irreversibility and generalized noise.
\newblock {\em Physical Review}, 83(1):34, 1951.

\bibitem{johnson1928thermal}
John~Bertrand Johnson.
\newblock Thermal agitation of electricity in conductors.
\newblock {\em Physical review}, 32(1):97, 1928.

\bibitem{nyquist1928thermal}
Harry Nyquist.
\newblock Thermal agitation of electric charge in conductors.
\newblock {\em Physical review}, 32(1):110, 1928.

\bibitem{rochman2016grow}
Nash Rochman, Fangwei Si, and Sean~X Sun.
\newblock To grow is not enough: impact of noise on cell environmental response
  and fitness.
\newblock {\em Integrative Biology}, 8(10):1030--1039, 2016.

\bibitem{verley2011modified}
Gatien Verley, K~Mallick, and D~Lacoste.
\newblock Modified fluctuation-dissipation theorem for non-equilibrium steady
  states and applications to molecular motors.
\newblock {\em EPL (Europhysics Letters)}, 93(1):10002, 2011.

\end{thebibliography}

\end{document}